# Alternative explanation for the steep subthreshold slope in ferroelectric FETs

Jan Van Houdt and Philippe Roussel
imec, Kapeldreef 75, B-3001 Leuven, Belgium
*) also with K.U. Leuven, Arenbergpark 10, B-3001 Leuven, Belgium
phone: +32 16 28 1268, vanhoudt@imec.be

*Abstract* — Since many years, sub-60mV/decade switching has been reported in ferroelectric FETs. However, thus far these reports have lacked full physical explanation since they typically use a negative capacitance in the ferroelectric layer to be able to explain the experimental observations. Because negative capacitance as such is not a physical concept, we propose an alternative model that relies on the non-linear and non-equilibrium behavior of the ferroelectric layer. It is shown that a steep subthreshold slope can be obtained by a 2-step switching process, referred to as nucleation and propagation. Making use of the concept of domain wall motion as known also from fracture dynamics, we are able to explain the steep slope effect. A simple mathematical model is added to further describe this phenomenon, and to further investigate its eventual benefit for obtaining steep slope transistors in the sub-10nm era.

## I. INTRODUCTION

Ferroelectrics have regained a lot of interest since the discovery of a ferroelectric phase in $HfO_2$ [1], the most obvious application being nonvolatile memories (as in the case of perovskites [2]). However, many research groups have reported steep subthreshold slope behavior of ferroelectric FETs as well. This has increased the expectations for overcoming the Boltzmann tyranny [3]. The full explanation for this phenomenon has however been lacking since. When considering the peculiar dynamics of ferroelectric switching we can obtain a steep slope without the need for introducing negative capacitance values. In order to obtain this result, we have to include the concept of domain wall motion which allows to include the dielectric response acceleration needed to amplify the surface potential. This theory is then formalized by means of a simple charge balance equation to describe the phenomenon in a more quantitative way. A *non-equilibrium dielectric response* is found as a main condition for showing steep slope in FeFETs.

## II. PHYSICAL MODEL

In contrast to paraelectric polarization, when switching a ferroelectric layer the polarization of the layer is highly non-linear as a function of the applied field. This is clear from the initial P(V) curve as measured on any ferroelectric capacitor when starting from unpolarized (zero) state (Fig. 1). This can be explained by the increasing dielectric response to the external field when increasing the voltage (super-linear behavior of the polarization or increasing dielectric *susceptibility*, the latter being equal to the slope of the P(E) curve -1). From basic electrostatics this can be explained by the more than proportional increase in the volumetric density of the number of dipole moments with field (i.e. by nucleation). This happens already at low (subthreshold) voltage because the layer consists of many domains with different coercive fields. In fact, single domain loops can be shifted along the V-axis depending on the local electric field from the surrounding dipoles as described by the Preisach model [4] (Fig.1).

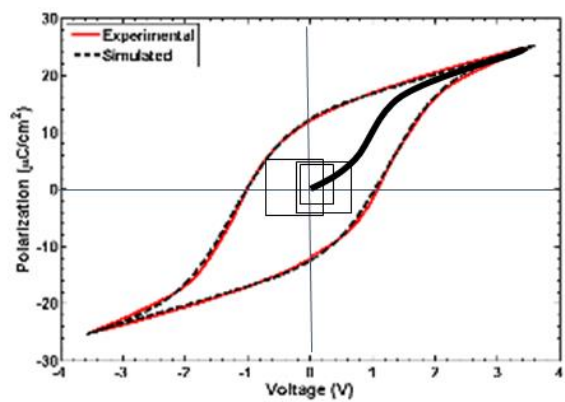

Fig 1 Typical P-V loop can be reconstructed from a large number of single domain (rectangular) loops as dictated by the Preisach model [4]. The S-shape curve initiating from the origin shows a superlinear behavior in the low voltage regime indicating the increased dielectric response already at low voltages.

As a result, the coercive voltage shows a wide distribution over the number of involved ferroelectric domains which are gradually starting to contribute to the overall polarization when the voltage increases. That explains the partial polarization starting already in the subthreshold regime and then accelerating with the externally applied field. From literature it is known that ferroelectric switching is typically accompanied also by domain wall motion [5]. This is usually requiring much less energy than the actual switching energy to reverse the dipoles by 180° [6]. This mechanism is also at the origin of crack propagation in materials: there is a critical strain at which the crack continues to grow with only a small external stimulus (positive feedback effect). This effect is correlated with twin motion which is known to happen at the speed of sound (or even faster, see [5] or [6] for the case of ferroelectric $BaTiO_3$). A similar twin motion is used to explain the domain wall motion in ferroelectric switching as well [6]. Since the speed of sound in such crystalline materials is on the order of 10km/s (or 10nm/ps) the propagation of the dipole switching is much faster than the typical gate pulse applied to the gate of a state-of-the-art transistor especially in a laboratory environment (>10ns). That means that the polarization (i.e. density of dipoles) is further *accelerated* as a function of time without following the externally applied voltage conditions. The domain wall motion is known to be stalled by defects (so-called pinning [6]) which explains why the ferroelectric will not reach full saturation at small voltages and depolarization can take place.

### III. MATHEMATICAL DESCRIPTION

Since the material properties are changing during ferroelectric switching, the concept of dielectric constant becomes highly unpractical. But it is clear when using average values for the dielectric response (in space), the charge balance equation is still valid. This lemma can be derived directly from the basic Maxwell equations (not shown here).

Splitting the FeFET structure into 2 capacitors, one being the ferroelectric part and the other one describing the series of the interfacial oxide capacitor and the substrate depletion capacitor (in subthreshold), we get the simple capacitive divider shown in Fig.2. However, both capacitors are voltage dependent. We can equalize the charge on both capacitors and obtain Eq. (1).

$$\frac{C_{ox}}{\sqrt{1 + \frac{2 C_{ox}^2 \cdot V_{int}}{q_e \cdot N_a \cdot \varepsilon_{Si}}}} \cdot V_{int} = \left(C_0 + \beta \cdot V_g^n\right) \cdot \left(V_g - V_{int}\right)$$

Eq.(1)

where $N_a$ is the doping concentration of the substrate and $\varepsilon_{Si}$ is the permittivity of silicon. The other parameters are clear from Fig.2. The bottom capacitor formula (left side of the equation) is taken from [7] which provides a closed formula for the depletion capacitor in series with the interfacial one, while for the ferroelectric capacitor we need to include an increasing function with time to take care of the propagation effect. Since we consider the case where $V_g = \alpha \cdot t$ this is equivalent to assuming that the ferroelectric capacitor is a function of $V_g$ only.

The solution of Eq.(1) for reasonable values of the parameters (see caption) yields the characteristics shown in Fig.3. The green (lower) curve is the physical solution of the charge balance equation.

If we raise the voltage at the gate, the voltage over both capacitors will increase, which implies that the bottom capacitor decreases and the ferroelectric one increases. Thus, the internal voltage $V_{int}$ will start to increase faster than at t=0. If the response of the material reaches the propagation phase (i.e. domain growth takes over), the polarization keeps on increasing to the point where the voltage over the ferroelectric capacitor starts to *decrease* with time. At that point, the internal voltage $V_{int}$ increases faster than the gate voltage and an amplification is seen at the level of the surface potential ($d\psi_s/dV_g > 1$).

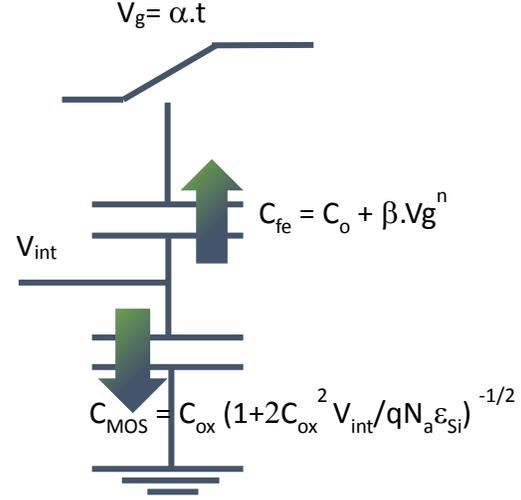

Fig 2 Capacitive divider consisting of 2 non-linear capacitors with a time dependent response added for the ferroelectric capacitor. $V_{int}$ is the potential at the interface between the interfacial and the ferroelectric layer or at the intermediate gate when considering MFMIS structures [3]. It can be shown that the non-linearity of the MOS capacitor is enhancing the 'boosting' of the surface potential.

In summary, it is possible to observe a boost or amplification of the surface potential response $d\psi_s/dV_g$. It suffices that the internal node goes up faster than the gate bias, which is the case when the dielectric response stops following the voltage over the capacitor and becomes a function of time only. At that point, the ferroelectric capacitor increases further with decreasing voltage (Fig.4).

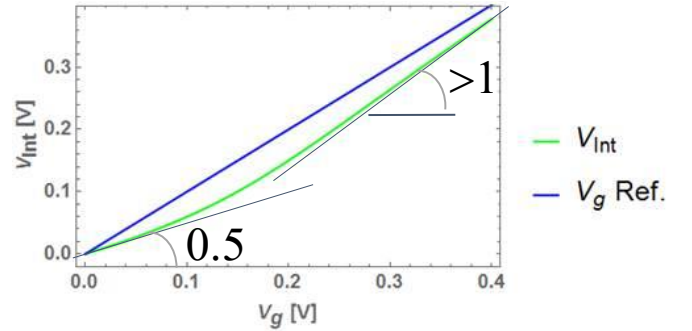

Fig 3 By taking $C_o = C_{ox}$ a capacitive division of exactly 0.5 is obtained at $V_g = t = 0$. Solving Eq.(1) for $V_{int}$ with $C_{ox} = C_o$, n=3, β=3E-4 shows an increasing slope for $V_{int}$ beyond that of $V_g$. This is causing a higher surface potential *response* than in the case without ferroelectric capacitor.

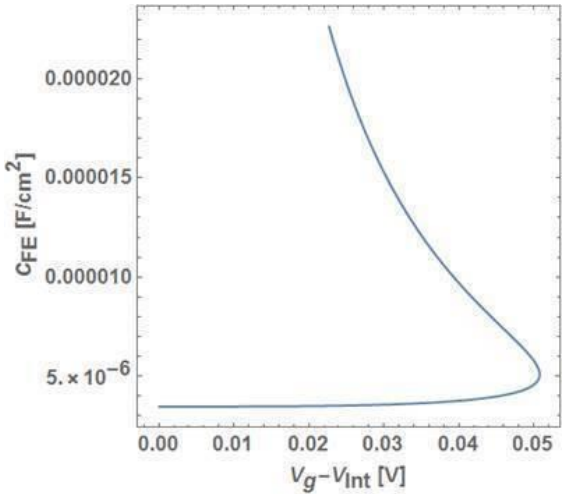

Fig 4 Dynamic value of the ferroelectric capacitor as calculated from Eq.(1).

## IV. Discussion

According to the model, the steep slope effect will only appear under certain conditions. As can be seen from Fig.3 it will not be there from $V_g=0$ but only when the voltage over the capacitor starts to decrease. It is, however, possible that the device reaches inversion faster than in the reference case provided that the nucleation of the ferroelectric domains happens faster than the gate ramp rate. Taking some typical numbers from [8] ($f_o=1$THz) and [9] (Landau barrier = 100meV for the case of orthorhombic $HfO_2$), one can calculate from an attempt-to-escape-frequency model that nucleation happens with a time constant on the order of 50ps at room temperature, which is of the order of state-of-the-art MOS transistors (~10ps). In practice, this time constant is reduced by the application of an electric field. Alternatively, ferroelectrics with larger $f_o$ or lower barrier may be required. The fact that steeper slope is obtained for larger hysteresis can be explained from the basic P-V loop: the more domains are switching, the stronger the effect will be and the larger the remnant polarization. At lower fields, the partial polarization is expected to disappear on a relatively short time scale due to depolarization (so-called subloop operation) [10].

Fig.5 summarizes how the steeper internal voltage is equivalent to a steeper surface potential.

## V. Conclusion

It has been shown that the domain wall motion effect in ferroelectrics can be described as a walk-out (or 'domino' effect) of the dielectric response in the ferroelectric capacitor which leads to decreasing voltage over the capacitor. As a consequence, the surface potential can increase faster than the external gate voltage leading to steep slope behavior in a limited range of gate biases. The presented model will be helpful in order to better understand the behavior of ferroelectric FETs in the subthreshold regime as well as its potential applications.

### Acknowledgment

This work is supported by imec's Industrial Affiliation Program on Ferroelectrics.

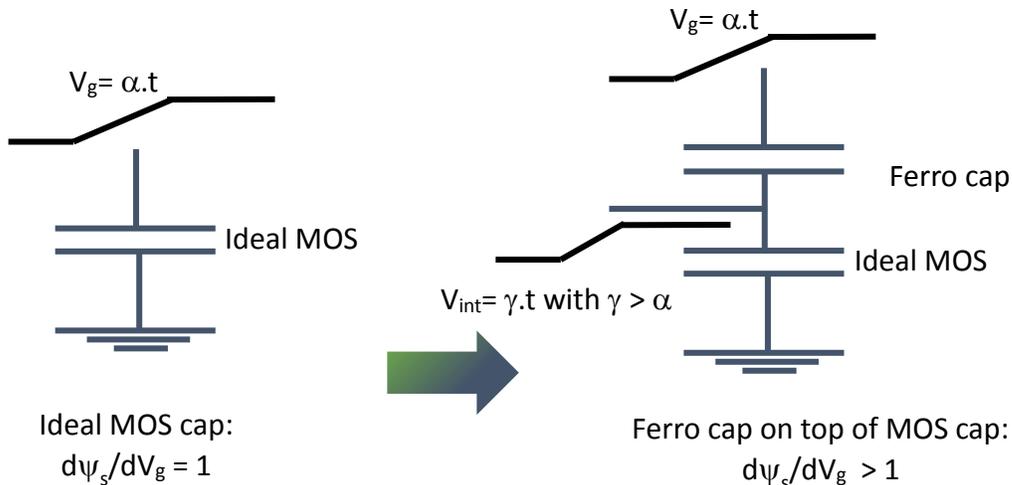

Fig 5 surface potential boosting from steeper internal voltage (equivalent to steeper subthreshold slope in a transistor using such a stack).